\documentclass[12pt]{article}
\usepackage{amsmath}
\usepackage{amsfonts}
\usepackage{amssymb}
\usepackage{srcltx}

\numberwithin{equation}{section}

\newcommand{\beq}{\begin{equation}}
\newcommand{\eeq}{\end{equation}}

\makeatletter
\newcommand{\rd}{\@ifnextchar^{\DIfF}{\DIfF^{}}}
\def\DIfF^#1{%
   \mathop{\mathrm{\mathstrut d}}%
   \nolimits^{#1}\gobblespace}
\def\gobblespace{\futurelet\diffarg\opspace}
\def\opspace{%
   \let\DiffSpace\!%
   \ifx\diffarg(%
   \let\DiffSpace\relax
   \else
   \ifx\diffarg[%
   \let\DiffSpace\relax
   \else
   \ifx\diffarg\{%
   \let\DiffSpace\relax
   \fi\fi\fi\DiffSpace}
\providecommand*{\dder}[3][]{%
\frac{\rd^{#1}#2}{\rd #3^{#1}}}
\providecommand*{\pder}[3][]{%
\frac{\partial^{#1}#2}{\partial #3^{#1}}}
\providecommand*{\iu}%
{\ensuremath{\mathrm{i}\,}}

\begin{document}
\noindent
{\Large {\bf B\"acklund transformations for certain rational solutions of Painlev\'e VI }}
\vskip 9 mm
\begin{center}
\begin{minipage}[t]{70mm}
{\bf Henrik Aratyn}\\
\\
Department of Physics,\\
University of Illinois at Chicago,\\
845 W. Taylor St.,\\
Chicago, IL 60607-7059\\
e-mail: aratyn@uic.edu\\
\end{minipage}
\begin{minipage}[t]{70mm}
{\bf Johan van de Leur}\\
\\
Mathematical Institute,\\
University of Utrecht,\\
P.O. Box 80010, 3508 TA Utrecht,\\
The Netherlands\\
e-mail: J.W.vandeLeur@uu.nl
\end{minipage}
\\
\end{center}
\begin{center}
\begin{minipage}[t]{140mm}
{\bf Abstract.}
We introduce certain B\"acklund transformations for  rational
solutions  of the Painlev\'e VI equation. These transformations act on
a family of Painlev\'e VI tau functions. They are obtained from
reducing the Hirota bilinear equations that describe the relation between certain points in the 3 component polynomial KP Grassmannian. In this way we obtain transformations that act on the root lattice of $A_5$. We also show that this $A_5$ root lattice can be related to the $F_4^{(1)}$ root lattice. We thus obtain B\"acklund transformations that relate Painlev\'e VI  tau functions, parametrized by the elements of this $F_4^{(1)}$ root lattice. \end{minipage}
\end{center}
\section{Introduction}
In  \cite{AL-IMRN}, which was a generalization of \cite{AL}( see also \cite{Lor}), we showed that there is a connection between certain homogeneous solutions of the $3$-component KP hierarchy and certain rational solutions (cf. \cite{Maz}) of the Painlev\'e VI equation:
\begin{equation}
\begin{split}
 \dder[2]{y}{t} &=\frac{1}{2}\left( 
\frac{1}{y} + \frac{1}{y-1} + \frac{1}{y-t} 
\right)\left(\dder{y}{t}\right)^2
-\left(\frac{1}{t} + \frac{1}{t-1} + \frac{1}{y-t} 
 \right)\dder{y}{t}\\
&+ \frac{y(y-1)(y-t)}{t^2(t-1)^2}
\left\{ \alpha + \beta \frac{t}{y^2}
+ \gamma \frac{t-1}{(y-1)^2}
+ \delta \frac{t(t-1)}{(y-t)^2} \right\}.
\end{split}
\label{eq:P6}
\end{equation}
In this publication we focus on the B\"acklund transformations for  the solutions of  \cite{AL-IMRN}. See also \cite{kakei} for connections of $gl_3$ KP hierarchy to Painlev\'e VI. Instead of obtaining  B\"acklund transformations for the Painlev\'e VI equation, we obtain such transformations for the so-called Jimbo-Miwa-Okamoto $\sigma$-form of the Painlev\'e VI
equation \cite{JM}:
\begin{equation}
\dder{\sigma}{t} \left(t (t-1) \dder[2]{\sigma}{t}\right)^2+
\left( \dder{\sigma}{t}  \left[2 \sigma
-(2t-1) \dder{\sigma}{t}\right] +v_1v_2v_3v_4\right)^2
=\prod_{k=1}^4 \left( \dder{\sigma}{t}+v_k^2\right)\, ,
\label{eq:jmo}
\end{equation}
where
\begin{equation}
v_1+v_2=\sqrt{-2\beta},\;\; v_1-v_2=\sqrt{2\gamma},\;\;
v_3+v_4+1=\sqrt{1-2\delta},\;\; v_3-v_4=\sqrt{2\alpha}\, .
\label{eq:via}
\end{equation}
This $\sigma$ is related via some choice of variables to the 3-component KP tau-function $T$ by
\[
\sigma(t)=t(t-1)\frac{d \log T}{dt}-a t - b\, ,
\]
for  certain constants $a,b$.
In this paper we show that there exists such a tau-function for certain elements
in the root lattice of $sl_6$:
\begin{equation}
\label{root lattice A5}
Q(A_5)=\{\underline\alpha =\sum_{i=1}^6 \alpha_i\underline\delta_i\, |\, \sum_{i=1}^6 \alpha_i=0\}\, ,
\end{equation}
where $(\underline\delta_i)_j=\delta_{ij}$ and where we choose
\begin{equation}
\label{v-in-alpha-mu0}
v_i=\frac{\alpha_1+\alpha_3}2+\alpha_{3+i}\ \ (i=1,2,3),\qquad v_4=\frac{\alpha_1-\alpha_3}2\, .
\end{equation}
The equations of the 3-component KP and modified 3-component KP produce B\"acklund transformations on the above tau-functions
\begin{equation}
\label{blabla0}
T_{\underline \alpha+\underline\delta_i-\underline\delta_k}\partial_j(T_{\underline\alpha})-T_{\underline\alpha}\partial_j
(T_{\underline\alpha+\underline\delta_i-\underline\delta_k})+n_j(\underline\alpha;i,k)T_{\underline\alpha} T_{\underline\alpha+\underline\delta_i-\underline\delta_k}=\epsilon_{ijk}\, T_{\underline\alpha+\underline\delta_i-\underline\delta_j}
T_{\underline\alpha+\underline\delta_j-\underline\delta_k}\, ,
\end{equation}
for  distinct $i,j,k$ with $1\le j\le3$ and $ 1\le i, k\le 6$. Here
\[
\partial_j=b_j(t)\frac{d}{dt},\qquad\mbox {and}\quad b_1(t)= t(t-1),\quad 
b_2(t)=t,\quad
b_3(t)=-t^2
\]
and
$
n_1(\underline\alpha;i,k)$ is a certain constant, which is given in 
(\ref{nnn}).
From this we deduce the following B\"acklund equation for the Jimbo-Miwa-Okamoto $\sigma$-function for  distinct $i,j,k$ with $1\le j\le3$ and $ 1\le i, k\le 6$:
\begin{equation}
\label{blabla40}
\begin{split}
\sigma_{\underline\alpha+\underline\delta_i-\underline\delta_j}(t)&+\sigma_{\underline\alpha+\underline\delta_j-\underline\delta_k}(t)-\sigma_{\underline\alpha+\underline\delta_i-\underline\delta_k}(t)-\sigma_{\underline\alpha}(t)=\\
&=G_{ijk}(\underline\alpha;t)
+t(t-1)\frac{d}{dt}
\log\left(\sigma_{\underline\alpha}(t)
-\sigma_{\underline\alpha+\underline\delta_i-\underline\delta_k}(t)+H_{ijk}(\underline\alpha;t) \right)\, .
\end{split}
\end{equation}
Here 
$G_{ijk}(\underline\alpha;t)$ and $H_{ijk}(\underline\alpha;t)$ are certain first order polynomials that can be determined explicitly, see (\ref{GH}).
 
\section{The polynomial Grassmannian and the (modified) 3-component KP-hierarchy}
The geometry behind the rational Painlev\'e VI solutions of \cite{AL-IMRN} is the infinite polynomial
 (3-component) Grassmannian. Let 
$
H=\{ \sum_i c_i \lambda^i\, |\, c_i\in\mathbb{C}^3,\ c_i=0\ \mbox{for } i<< 0\}\, 
$
and 
$
H_+=\{ \sum_i c_i \lambda^i\, |\, c_i\in\mathbb{C}^3,\ c_i=0\ \mbox{for } i<0\}\, .
$
On $H$ we have a natural bilinear form given by 
\begin{equation}
\label{bil2}
(\sum_i c_i \lambda^i| \sum_j d_i \lambda^i)=\sum_i (c_i,d_{-1-i})\, ,
\end{equation}
where $(\cdot,\cdot)$ is the standard bilinear form on $\mathbb{C}^3$ given by
\begin{equation}
\label{bil1}
(w,v)=w_1v_1+w_2v_2+w_3v_3\, .
\end{equation}
The Grassmannian consists of linear subspaces $W\subset H$, that satisfy certain conditions.
Here we will consider only very special linear subspaces $W$ of $H$, viz. the ones that satisfy
the following conditions:
\begin{itemize}
\item There exist positive integers $m$ and $n$ such that
$
\lambda^{n} H_+\subset W\subset \lambda^{-m} H_+\, ,
$
\item $W$  satisfy the condition
$
\lambda W\subset W
$,
\item $W$ has a basis of elements $v(\lambda)$ that are homogeneous in $\lambda$, i.e. $\lambda\frac{d v(\lambda)}{d\lambda}=dv(\lambda)$ with $d\in\mathbb{Z}$.
\end{itemize}
All this gives that such a $W$ can be described as follows, see \cite{AL-IMRN} for more details.
Choose 3 linearly independent vectors in $\mathbb{C}^3$
\[
w^{(i)}=\left(w_1^{(i)},w_2^{(i)},w_3^{(i)}\right),\qquad i=1,2,3,
\]
and let 
\[
w_{(i)}=\left(w_{(i)}^1,w_{(i)}^2,w_{(i)}^3\right)
\]
be the dual basis with respect to the bilinear form (\ref{bil1}).

Let 
\begin{equation}
\label{mu}
\underline \mu=(\mu_1,\mu_2,\mu_3)=\mu_1\underline\epsilon_1+\mu_2\underline\epsilon_2+\mu_3\underline\epsilon_3\in\mathbb{Z}^3\, ,
\end{equation}
where $\underline\epsilon_j$ is a basis vector in $\mathbb{Z}^3$, so
$
\underline\epsilon_j=(\delta_{j1},\delta_{j2},\delta_{j3})$.
Then such a $W$ is equal to $W(\underline\mu)$, where
\[
W(\underline\mu)=\sum_{i\ge \mu_1}\mathbb{C}\lambda^iw^{(1)}+\sum_{j\ge\mu_2}\mathbb{C}\lambda^jw^{(2)}+\sum_{k\ge\mu_3}\mathbb{C}\lambda^kw^{(3)}\, .
\]
Let $e_a$, $a=1,2,3$, be the standard basis of $\mathbb{C}^3$, then
\[
W(\underline\mu)=\sum_{i= \mu_1}^{\max \mu_\ell-1}\mathbb{C}\lambda^iw^{(1)}+\sum_{j=\mu_2}^{\max \mu_\ell-1}\mathbb{C}\lambda^jw^{(2)}+\sum_{k=\mu_3}^{\max \mu_\ell-1}\mathbb{C}\lambda^kw^{(3)}+\sum_{a=1}^3\sum_{m\ge \max \mu_\ell}\mathbb{C}\lambda^me_a\, .
\]
Note that
\[
W(\underline 0)=
\sum_{a=1}^3\sum_{m\ge 0}\mathbb{C}\lambda^me_a=H_+\, .
\]
With respect to the bilinear form (\ref{bil2}) on $H$ we can find the maximal orthocomplement $W^\perp(\underline\mu)$.
This space is given by
\[
W^\perp(\underline\mu)=\sum_{i\ge -\mu_1}\mathbb{C}\lambda^iw_{(1)}+\sum_{j\ge -\mu_2}\mathbb{C}\lambda^jw_{(2)}+\sum_{k\ge -\mu_3}\mathbb{C}\lambda^kw_{(3)}\, .
\]
Note that
\[
W^\perp(\underline\mu)=\sum_{i= -\mu_1}^{\max -\mu_\ell-1}\mathbb{C}\lambda^iw_{(1)}+\sum_{j=\mu_2}^{\max -\mu_\ell-1}\mathbb{C}\lambda^jw_{(2)}+\sum_{k=\mu_3}^{\max -\mu_\ell-1}\mathbb{C}\lambda^kw_{(3)}+\sum_{a=1}^3\sum_{m\ge \max -\mu_\ell}\mathbb{C}\lambda^me_a\, .
\]
If we define the following ordering on $\mathbb{Z}^3$
\[
\underline \mu\le\underline\lambda\quad\mbox{if }\mu_i\le \lambda_i\quad\mbox{for all }i=1,2,3\, ,
\]
then
\[
W(\underline \lambda)\subset
W(\underline \mu)\quad\mbox{and }W^\perp(\underline \mu)\subset W^\perp(\underline \lambda)\quad\mbox{iff } \underline\mu\le\underline\lambda.
\]
Next, we associate to $W(\underline\mu)$ the following vector in a semi-infinite wedge space:
\begin{equation}
\label{vect}
\begin{split}
|W(\underline\mu)\rangle=&\lambda^{\mu_1}w^{(1)}\wedge
\lambda^{\mu_1+1}w^{(1)}\wedge
\cdots\wedge
\lambda^{\max \mu_\ell-1}w^{(1)}\wedge
\lambda^{\mu_2}w^{(2)}\wedge
\lambda^{\mu_2+1}w^{(2)}\wedge
\cdots\\[2mm]
&\qquad\cdots\wedge\lambda^{\max \mu_\ell-1}w^{(2)}\wedge
\lambda^{\mu_3}w^{(3)}\wedge
\lambda^{\mu_3+1}w^{(3)}\wedge
\cdots\wedge
\lambda^{\max \mu_\ell-1}w^{(3)}\wedge\\[2mm]
&\qquad\lambda^{\max \mu_\ell}e_1\wedge
\lambda^{\max \mu_\ell}e_2\wedge
\lambda^{\max \mu_\ell}e_3\wedge
\lambda^{\max \mu_\ell+1}e_1\wedge
\lambda^{\max \mu_\ell+1}e_2\wedge\cdots\, .
\end{split}
\end{equation}
If we define the grading
\[
\deg( |W(\underline 0)\rangle)=0\quad \mbox{and}\quad \deg(\lambda^kw^{(j)})=\frac12-k\, ,
\]
then
\begin{equation}
\label{deg1}
\deg( |W(\underline\mu)\rangle)=\frac12\left(\mu_1^2+\mu_2^2+\mu_3^2\right)\, .
\end{equation}
For any $v\in \left(\mathbb{C}[\lambda,\lambda^{-1}]\right)^3$ we can define creation and annihilation operators, see e.g. \cite{KL} for more details.
Let $v_0\wedge v_1\wedge v_2 \wedge \cdots $ be an element in the semi-infinite wedge space, then we define
\[
\psi^+(v)v_0\wedge v_1\wedge v_2 \wedge \cdots=
 v\wedge v_0\wedge v_1\wedge v_2 \wedge \cdots
\]
and
\[ 
\psi^-(v) v_0\wedge v_1\wedge v_2 \wedge \cdots=
 \sum_{i=0}^\infty (-)^i(v |v_i) v_0\wedge\cdots \wedge v_{i-1}\wedge  v_{i+1}\wedge \cdots\, .
\]
These elements form a Clifford algebra, they satisfy the anti-commutation relations
\[
\begin{split}
\psi^+(v)\psi^+(w)+\psi^+(v)\psi^+(w)=0\, ,&\qquad
\psi^-(v)\psi^-(w)+\psi^-(v)\psi^-(w)=0\, ,\\[2mm]
\psi^+(v)\psi^-(w)+\psi^+(v)\psi^-(w)=(v|w)\, .
\end{split}
\]
Note that
\[
\psi^+(v)|W(\underline\mu)\rangle=0\quad \mbox{for } v\in W(\underline\mu),\qquad \psi^-(v)|W(\underline\mu)\rangle=0\quad \mbox{for } v\in W^\perp (\underline\mu)\, .
\]
Let $V_0= v_0\wedge v_1\wedge v_2 \wedge \cdots$ and 
$V_k=v_{-k}\wedge v_{-k+1}\wedge v_{-k+2} \wedge \cdots$ for $k\ge 0$, then, since 
\[
v=\sum_{a=1}^3\sum_{j\in\mathbb{Z}}(\lambda^{-j-1}e^a |v)\lambda^je_a\,  ,
\] we find that
\[
\begin{split}
\sum_{a=1}^3&\sum_{j\in\mathbb{Z}} \psi^+(\lambda^je_a)V_k
\otimes \psi^-(\lambda^{-j-1} e_a)V_0=\\[2mm]
=&\sum_{a=1}^3\sum_{j\in\mathbb{Z}}\lambda^je_a\wedge V_k\otimes
\left(\sum_{i=0}^\infty (-)^i(\lambda^{-j-1}e^a |v_i) v_0\wedge\cdots \wedge v_{i-1}\wedge  v_{i+1}\wedge \cdots\right)\\[2mm]
=&\sum_{a=1}^3\sum_{j\in\mathbb{Z}}\sum_{i=0}^\infty (-)^i(\lambda^{-j-1}e^a |v_i)\lambda^je_a\wedge V_k\otimes
 v_0\wedge\cdots \wedge v_{i-1}\wedge  v_{i+1}\wedge \cdots
\\[2mm]
=&\sum_{i=0}^\infty (-)^i v_i\wedge V_k\otimes v_0\wedge\cdots \wedge v_{i-1}\wedge  v_{i+1}\wedge \cdots\\[2mm]
=&\sum_{i=0}^\infty \psi^+(v_i) V_k\otimes \psi^-(v_i^*)V_0 =0\, .
\end{split}
\]
Here $v_i^*$ is the dual vector of $v_i$ with respect to the bilinear form (\ref{bil2}).
So in particular for $W(\underline \nu)\subset W(\underline \mu)$ one has
\begin{equation}
\label{bilequ}
\sum_{a=1}^3\sum_{k\in\mathbb{Z}} \psi^+(\lambda^ke_a)|W(\underline \mu)\rangle
\otimes \psi^-(\lambda^{-k-1} e_a)|W(\underline \nu)\rangle=0\quad \mbox{for }\underline \mu \le \underline \nu\, .
\end{equation}
In a similar way we see that for $i\ne j$:
\begin{equation}
\label{bil2equ}
\sum_{a=1}^3\sum_{k\in\mathbb{Z}} \psi^+(\lambda^ke_a)|W(\underline \mu+\underline\epsilon_i-\underline\epsilon_j)\rangle
\otimes \psi^-(\lambda^{-k-1} e_a)|W(\underline \mu)\rangle
=\epsilon{ij}|W(\underline \mu-\underline\epsilon_j)\rangle\otimes|W(\underline \mu+\underline\epsilon_i)\rangle
\, .
\end{equation}

Let $\delta(z-\lambda)=z^{-1}\sum_{n\in\mathbb{Z}}\left(\frac{z}{\lambda}\right)^n$ and introduce the fields
\[
\psi^\pm(\delta(z-\lambda)e_a)=\sum_{\in\mathbb{Z}}\psi^\pm(\lambda^{n}e_a)z^{-n-1}\, .
\]
Then (\ref{bilequ}) is equivalent to
\begin{equation}
\label{bilequ2}
\mbox{Res}_z\sum_{a=1}^3 \psi^+(\delta(z-\lambda)e_a)|W(\underline \mu)\rangle
\otimes \psi^-(\delta(z-\lambda)e_a)|W(\underline \nu)\rangle=0\quad \mbox{for }\underline \mu \le \underline \nu \,  .
\end{equation}

Using the boson-fermion correspondence we can express every such semi-infinite wedge $|W(\underline \mu)\rangle$ as a function in $F=\mathbb{C}[q_a,q_a^{-1},x_i^{(a)} ; a=1,2,3, i=1,2,3,\ldots]$. We identify $|W(\underline 0)\rangle$ with $1\in F$. Let $\sigma$ be the corresponding isomorphism,
then
\begin{equation}
\label{vertex}
\sigma\psi^\pm(\delta(z-\lambda)e_a)\sigma^{-1}=
q_a^{\pm 1} z^{\pm q_a\frac{\partial}{\partial q_a}}
\exp\left(\pm \sum_{i=1}^\infty x^{(a)}_iz^i\right)
\exp\left(\mp \sum_{i=1}^\infty \frac{\partial}{\partial x^{(a)}_i}\frac{z^{-i}}{i}\right)\, .
\end{equation}
Unfortunately the $q_a$ and $q_b$ for $a\ne b$ anticommute,
which means that we have to order them. We assume that 
\begin{equation}
\label{tau1}
\sigma(|W(\underline 0)\rangle)=1\qquad\mbox{and }
\sigma(|W(\underline \mu)\rangle)=\sum_{\underline \alpha\in\mathbb{Z}^3} \tau_{\underline \alpha}(\underline \mu;x)q_1^{\alpha_1}q_2^{\alpha_2}q_3^{\alpha_3}\, .
\end{equation}
It is straightforward to check that
\begin{equation}
\label{sl6}
\tau_{\underline \alpha}(\underline \mu;x)=0\quad\mbox{for }
\mu_1+\mu_2+\mu_3+\alpha_1+\alpha_2+\alpha_3\ne 0\, 
\end{equation}
and using (\ref{deg1}) that
\begin{equation}
\label{deg2}
R(\underline \mu,\underline \alpha):=\deg\left(\tau_{\underline \alpha}(\underline \mu;x)\right)=\frac12 \left(\mu_1^2+\mu_2^2+\mu_3^2-\alpha_1^2-\alpha_2^2-\alpha_3^2\right)\, .
\end{equation}
Having this in mind, it will be useful to introduce the following subset of $\mathbb{Z}^3$:
\[
L_{\underline\mu}=\{\underline\alpha\in \mathbb{Z}^3\, |\, 
\mu_1+\mu_2+\mu_3+\alpha_1+\alpha_2+\alpha_3=0\}\, .
\]
Note that since the form of the vectors $|W(\underline\mu)\rangle$ and $|W(\underline\mu-(\underline\epsilon_1+\underline\epsilon_2+\underline\epsilon_3))\rangle$
(\ref{vect}) are similar, one finds that 
\begin{equation}
\label{equaltau}
\tau_{\underline \alpha}(\underline \mu;x)=(-1)^{\alpha_2}
\tau_{\underline \alpha+(\underline\epsilon_1+\underline\epsilon_2+\underline\epsilon_3)}(\underline \mu-(\underline\epsilon_1+\underline\epsilon_2+\underline\epsilon_3);x)\, .
\end{equation}

We can use (\ref{vertex}) and (\ref{tau1}) to rewrite
(\ref{bilequ2}) as a generating series of Hirota bilinear equations. We forget the tensor symbol and write $x'$ for its first component and $x''$ for its second component.
Define
\[
q^{\underline\alpha}=q_1^{\alpha_1}q_2^{\alpha_2}q_3^{\alpha_3}
\]
and let 
\[
\varepsilon (\underline\epsilon_{j},\underline\alpha)
=\begin{cases}
1&\mbox{for } j=1,\\
(-1)^{\alpha_1}&\mbox{for } j=2,\\
(-1)^{\alpha_1+\alpha_2}&\mbox{for } j=3,
\end{cases}
\]
then (\ref{bilequ2}) is equivalent to
\begin{equation}
\label{bilequ3}
\aligned
\text{Res}_z &(
\sum^{3}_{a=1} \sum_{\underline\alpha\in L_{\underline\mu} ,\underline\beta
\in L_{\underline\nu}} \varepsilon (\underline\epsilon_{a},\underline\alpha - \underline\beta)
z^{\alpha_a -\beta_a}  \exp (\sum^{\infty}_{k=1}
(x^{(a)^{\prime}_{k}} - x^{(a)^{\prime \prime}}_{k} )z^{k}) 
\\
& \exp
( -\sum^{\infty}_{k=1} ( \frac{\partial}{\partial
x^{(a)^{\prime}}_{k}} -
\frac{\partial}{\partial x^{(a)^{\prime \prime}}_{k}} ) \frac{z^{-k}}{k}
) \tau_{\underline\alpha}(\underline\mu;x^{\prime})(q^{\underline\alpha + \underline\epsilon_{a}})^{\prime}
\tau_{\underline\beta}(\underline \nu ;x^{\prime \prime})(q^{\underline\beta - \underline\epsilon_{a}})^{\prime
\prime}) = 0,  \quad\underline\mu\le \underline\nu
\endaligned 
\end{equation}
and
(\ref{bil2equ}) is equivalent to ( $\epsilon_{ij}=\varepsilon (\underline\epsilon_{i},\underline\epsilon_j)$):
\begin{equation}
\label{bil2equ3}
\aligned
\text{Res}_z &(
\sum^{3}_{a=1} \sum_{\underline\alpha\in L_{\underline\mu+\underline\epsilon_i-\underline\epsilon_j} ,\underline\beta
\in L_{\underline\mu}} \varepsilon (\underline\epsilon_{a},\underline\alpha - \underline\beta)
z^{\alpha_a -\beta_a}  \exp (\sum^{\infty}_{k=1}
(x^{(a)^{\prime}}_{k} - x^{(a)^{\prime \prime}}_{a} )z^{k}) 
\\
& \exp
( -\sum^{\infty}_{k=1} ( \frac{\partial}{\partial
x^{(a)^{\prime}}_{k}} -
\frac{\partial}{\partial x^{(a)^{\prime \prime}}_{k}} ) \frac{z^{-k}}{k}
) \tau_{\underline\alpha}(\underline\mu +\underline\epsilon_i-\underline\epsilon_j ;x^{\prime})(q^{\underline\alpha + \underline\epsilon_{a}})^{\prime}
\tau_{\underline\beta}(\underline \mu ;x^{\prime \prime})(q^{\underline\beta - \underline\epsilon_{a}})^{\prime
\prime}) \\=&  \epsilon_{ij}\sum_{\underline\gamma\in L_{\underline\mu-\underline\epsilon_j} ,\underline\delta
\in L_{\underline\mu+\underline\epsilon_i}} 
\tau_{\underline\gamma}(\underline\mu -\underline\epsilon_j ;x^{\prime})(q^{\underline\gamma} )^{\prime}
\tau_{\underline\delta}(\underline \mu +\underline\epsilon_i;x^{\prime \prime})(q^{\underline\delta })^{\prime
\prime})\, .
\endaligned 
\end{equation}
Taking the coefficient of $(q^{\underline\alpha})^\prime (q^{\underline\beta})^{\prime \prime}$ in ({\ref{bilequ3}) for $\underline\alpha\in L_{\underline\mu-\underline\epsilon_i}$ and $\underline\beta\in L_{\underline\nu+\underline\epsilon_i}$
we obtain:
\begin{equation}
\label{bilequ4}
\aligned
\text{Res}_z &(
\sum^{3}_{a=1}  \varepsilon (\underline\epsilon_{a},\underline\alpha - \underline\beta)
z^{\alpha_a -\beta_a-2}  \exp (\sum^{\infty}_{k=1}
(x^{(a)^{\prime}}_{k} - x^{(a)^{\prime \prime}}_{k} )z^{k}) 
\\
& \exp
( -\sum^{\infty}_{k=1} ( \frac{\partial}{\partial
x^{(a)^{\prime}}_{k}} -
\frac{\partial}{\partial x^{(a)^{\prime \prime}}_{k}} ) \frac{z^{-k}}{k}
) \tau_{\underline\alpha-\underline\epsilon_a}(\underline\mu;x^{\prime})
\tau_{\underline\beta+\underline\epsilon_a}(\underline \nu ;x^{\prime \prime}) = 0,
\quad\underline\mu\le \underline\nu
\endaligned 
\end{equation}
and in a similar way ({\ref{bil2equ3}) gives:
\begin{equation}
\label{bil2equ4}
\aligned
\text{Res}_z &(
\sum^{3}_{a=1}  \varepsilon (\underline\epsilon_{a},\underline\alpha - \underline\beta)
z^{\alpha_a -\beta_a-2}  \exp (\sum^{\infty}_{k=1}
(x^{(a)^{\prime}}_{k} - x^{(a)^{\prime \prime}}_{k} )z^{k}) 
\\
& \exp
( -\sum^{\infty}_{k=1} ( \frac{\partial}{\partial
x^{(a)^{\prime}}_{k}} -
\frac{\partial}{\partial x^{(a)^{\prime \prime}}_{k}} ) \frac{z^{-k}}{k}
) \tau_{\underline\alpha-\underline\epsilon_a}(\underline\mu+\underline\epsilon_i-\underline\epsilon_j;x^{\prime})
\tau_{\underline\beta+\underline\epsilon_a}(\underline \mu ;x^{\prime \prime}) \\
= &\epsilon_{ij}
\tau_{\underline\alpha}(\underline\mu-\underline\epsilon_j;x^{\prime})
\tau_{\underline\beta}(\underline \mu +\underline\epsilon_i;x^{\prime \prime})  .
\endaligned 
\end{equation}
Now making the change of variables
$x^{(j)}_{k} = { \frac{1}{2}} (u^{(j)^{\prime}}_{k} + u^{(j)^{\prime
\prime}}_{k})$, $ y^{(j)}_{k} = { \frac{1}{2}} (u^{(j)^{\prime}}_{k} -
u^{(j)^{\prime \prime}}_{n})$, one gets for (\ref{bilequ4}  ) for $\underline\mu\le \underline\nu$:
\begin{equation}
\label{bilequ5}
\aligned
\text{Res}_z &(
\sum^{3}_{j=1}  \varepsilon (\underline\epsilon_{j},\underline\alpha - \underline\beta)
z^{\alpha_j -\beta_j-2}
  \\
& \times \exp (\sum^{\infty}_{k=1} 2y^{(j)}_{k} z^{k}) \exp
(-\sum^{\infty}_{k=1}
\frac{\partial}{\partial y^{(j)}_{k}} \frac{z^{-k}}{k}) \tau_{\underline\alpha-\underline\epsilon_j}(\underline\mu; x+y) \tau_{\underline\beta+\underline\epsilon_j}(\underline \nu ;x-y)) = 0\, .
\endaligned 
\end{equation}
Using elementary Schur functions we rewrite this again as
\begin{equation}
\label{bilequ6}
\sum^{3}_{j=1} \varepsilon (\underline\epsilon_{j},\underline\alpha - \underline\beta)
 \sum^{\infty}_{k=0} S_{k}(2y^{(j)})S_{k-1+\alpha_j
-\beta_j } (-\frac{\tilde \partial}{\partial y^{(j)}})
 \tau_{\underline\alpha-\underline\epsilon_j}(\underline\mu; x+y) \tau_{\underline\beta+\underline\epsilon_j}(\underline \nu ;x-y)) = 0.
\end{equation}
Here and further we use the notation
$\frac{\tilde \partial}{\partial y} = (\frac{\partial}{\partial y_{1}} ,
\frac{1}{2} \frac{\partial}{\partial  y_{2}},\ \frac{1}{3}
\frac{\partial}{\partial y_{3}}, \ldots )$.
Using Taylor's formula we can rewrite this once more:
\begin{equation}
\label{bilequ7}\aligned
\sum^{3}_{j=1} &\varepsilon (\underline\epsilon_{j},\underline\alpha - \underline\beta)
 \sum^{\infty}_{k=0} S_{k}(2y^{(j)})S_{k-1+\alpha_j
-\beta_j } (-\frac{\tilde \partial}{\partial t^{(j)}}) \\
& \times e^{\sum^{3}_{j=1}
\sum^{\infty}_{r=1}y^{(j)}_{r}\frac{\partial}{\partial
t^{(j)}_{r}} } \tau_{\underline\alpha-\underline\epsilon_j}(\underline\mu;x+t)\tau_{\underline\beta+\underline\epsilon_j}(\underline \nu ;x-t)|_{t=0} = 0.
\endaligned 
\end{equation}
This last equation can be written as the following generating series of Hirota
bilinear equations:
\begin{equation}
\label{bilequ88}\sum^{3}_{j=1} \varepsilon (\underline\epsilon_{j},\underline\alpha - \underline\beta)
 \sum^{\infty}_{k=0} S_{k}(2y^{(j)})S_{k-1+\alpha_j
-\beta_j } (-\widetilde{D^{(j)}})  e^{\sum^{3}_{j=1} \sum^{\infty}_{r=1} y^{(j)}_{r}D^{(j)}_{r}}
\tau_{\underline\alpha-\underline\epsilon_j}(\underline\mu)\cdot\tau_{\underline\beta+\underline\epsilon_j}(\underline \nu ) = 0\, ,
\end{equation}
for all $\underline\alpha \in L_{\underline\mu-\underline\epsilon_i}$, $\underline\beta \in L_{\underline\nu+\underline\epsilon_i}$ and $\underline\mu\le \underline\nu$, see \cite{KL} for more details.

Now take $\underline\mu=\underline\nu$, then for $\underline\alpha \in L_{\underline\mu}$
and $1\le i,j\le 3$
distinct indices $i$ and $j$ one finds the following equation:
\begin{equation}
\label{eq1}
D^{(i)}_{1} D^{(j)}_{1} \tau_{\underline\alpha}(\underline\mu) \cdot \tau_{\underline\alpha}(\underline\mu) = 2
\tau_{\underline\alpha+\underline\epsilon_i-\underline\epsilon_j}(\underline\mu) \tau_{\underline\alpha+\underline\epsilon_j-\underline\epsilon_i} (\underline\mu)
\end{equation}
and
for each 
triple of distinct indices
$i,j,k$:
\begin{equation}
\label{eq2}
D^{(j)}_{1} \tau_{\underline\alpha}(\underline\mu)\cdot \tau_{\underline\alpha+\underline\epsilon_i-\underline\epsilon_k}(\underline\mu) =
\varepsilon_{ijk} 
\tau_{\underline\alpha+\underline\epsilon_i-\underline\epsilon_j}(\underline\mu) \tau_{\underline\alpha+\underline\epsilon_j-\underline\epsilon_k}(\underline\mu)\, .
\end{equation}

If $\underline\mu=\underline\nu-\underline\epsilon_\ell$, choose first $\underline\alpha$ and $\underline\beta$ such that  $\underline\alpha-\underline\beta=\underline\epsilon_1+\underline\epsilon_2+\underline\epsilon_3$, then we find the following Hirota-Miwa equation:
\begin{equation}
\label{Miwa1}
\tau_{\underline\beta+\underline\epsilon_2+\underline\epsilon_3}(\underline\mu)\tau_{\underline\beta+\underline\epsilon_1}(\underline\mu+\underline\epsilon_\ell)-
\tau_{\underline\beta+\underline\epsilon_1+\underline\epsilon_3}(\underline\mu)\tau_{\underline\beta+\underline\epsilon_2}(\underline\mu+\underline\epsilon_\ell)
+
\tau_{\underline\beta+\underline\epsilon_1+\underline\epsilon_2}(\underline\mu)\tau_{\underline\beta+\underline\epsilon_3}(\underline\mu+\underline\epsilon_\ell)
=0\, .
\end{equation}
Secondly choose $\underline\alpha$ and $\underline\beta$ such that  $\underline\alpha-\underline\beta=2\underline\epsilon_i+\underline\epsilon_j$, with $i$ and $j$ distinct, then we find
\begin{equation}
\label{eq3}
D_1^{(i)}\tau_{\underline\gamma}(\underline\mu)\cdot \tau_{\underline\gamma-\underline\epsilon_j}(\underline\mu+\underline\epsilon_\ell)=\epsilon_{ij}\tau_{\underline\gamma-\underline\epsilon_i}(\underline\mu+\underline\epsilon_\ell) \tau_{\underline\gamma+\underline\epsilon_i-\epsilon_j}(\underline\mu)\, ,
\end{equation}
or equivalently
\begin{equation}
\label{2eq3}
D_1^{(i)} \tau_{\underline\gamma}(\underline\mu)\cdot
\tau_{\underline\gamma+\underline\epsilon_j}(\underline\mu-\underline\epsilon_\ell)
=\epsilon_{ji}\tau_{\underline\gamma+\underline\epsilon_j-\underline\epsilon_i}(\underline\mu) \tau_{\underline\gamma+\underline\epsilon_i}(\underline\mu-\underline\epsilon_\ell)\, .
\end{equation}

In a similar way (\ref{bil2equ4}) can be rewritten as the following generating series of Hirota bilinear equations ($i\ne j$):
\begin{equation}
\label{bilequ8}
\begin{split}
\sum^{3}_{a=1}& \varepsilon (\underline\epsilon_{a},\underline\alpha - \underline\beta)
 \sum^{\infty}_{k=0} S_{k}(2y^{(a)})S_{k-1+\alpha_a
-\beta_a } (-\widetilde{D^{(a)}})  e^{\sum^{3}_{a=1} \sum^{\infty}_{r=1} y^{(a)}_{r}D^{(a)}_{r}}
\tau_{\underline\alpha-\underline\epsilon_a}(\underline\mu+\underline\epsilon_i-\underline\epsilon_j)\cdot\tau_{\underline\beta+\underline\epsilon_a}(\underline \mu )\\ =&\epsilon_{ij}\,
e^{\sum^{3}_{a=1} \sum^{\infty}_{r=1} y^{(a)}_{r}D^{(a)}_{r}}\tau_{\underline\alpha}(\underline\mu-\underline\epsilon_j)\cdot\tau_{\underline\beta}(\underline \mu +\underline\epsilon_i)\, ,
\end{split}
\end{equation}
for all $\underline\alpha \in L_{\underline\mu-\underline\epsilon_i}$, $\underline\beta \in L_{\underline\nu+\underline\epsilon_i}$ and $\underline\mu\le \underline\nu$.
Now taking $\underline\alpha-\underline\beta=\underline\epsilon_k+\underline\epsilon_\ell$, with $k\ne \ell$, where $k$  or $\ell$ may be equal to $i$ or $j$, we find another version of the Hirota-Miwa equation ($i\ne j$, $k\ne \ell$):
\begin{equation}
\label{Miwa2}
\epsilon_{k\ell}\tau_{\underline\beta+\underline\epsilon_\ell}(\underline\mu+\underline\epsilon_i-\underline\epsilon_j)\tau_{\underline\beta+\underline\epsilon_k}(\underline\mu)
+\epsilon_{\ell k}\tau_{\underline\beta+\underline\epsilon_k}(\underline\mu+\underline\epsilon_i-\underline\epsilon_j)\tau_{\underline\beta+\underline\epsilon_\ell}(\underline\mu)
-\epsilon_{ij}\tau_{\underline\beta+\underline\epsilon_k+\underline\epsilon_\ell}(\underline\mu-\underline\epsilon_j)\tau_{\underline\beta}(\underline\mu+\underline\epsilon_i)=0\, .
\end{equation}
Next taking $\underline\alpha-\underline\beta=2\underline\epsilon_k$, where $k$ may be equal to $i$ or $j$, we find ($i\ne j$):
\begin{equation}
\label{eq4}
D_1^{(k)}\tau_{\underline\gamma}(\underline\mu)\cdot \tau_{\underline\gamma}(\underline\mu+\underline\epsilon_i-\underline\epsilon_j)=\epsilon_{ji}\tau_{\underline\gamma+\underline\epsilon_k}(\underline\mu-\underline\epsilon_j) \tau_{\underline\gamma-\underline\epsilon_k}(\underline\mu+\underline\epsilon_i)\, .
\end{equation}
\vskip 20pt

In the above construction the pair 
\[
(\underline\alpha,\underline\mu)=(\alpha_1,\alpha_2,\alpha_3,\mu_1,\mu_2,\mu_3)=(\alpha_1,\alpha_2,\alpha_3,\alpha_4,\alpha_5,\alpha_6)
\] can be seen as an element in the root lattice $Q(A_5)$ of 
$sl_6$ (see(\ref{root lattice A5})). Note that  the tau function corresponding to such a pair  $(\underline\alpha,\underline\mu)$ is 0, whenever this pair is not  in $Q(A_5)$,  see (\ref{sl6}). A basis of this root lattice is given by the roots $\underline\delta_i-\underline\delta_{i+1}$ for $1\le i\le 5$.
Using the degree of the tau function given in (\ref{deg2}), we define a similar grading on this root lattice by
\begin{equation}
\label{deg3}
R(\underline\alpha)=R\left(\sum_{i=1}^6 \alpha_i\underline\delta_i\right)=\deg (\sum_{i=1}^6 \alpha_i\underline\delta_i)= \frac12\left(\alpha_4^2+\alpha_5^2+\alpha_6^2-\alpha_1^2-\alpha_2^2-\alpha_3^2\right)\, .
\end{equation}

In this light the equations (\ref{eq2}), (\ref{eq3}), (\ref{eq4}) can be rewritten to one equation. Let $\underline \beta$ be an element in the root lattice of $sl_6$, then for for  distinct $i,j,k$ with $1\le j\le3$ and $ 1\le i, k\le 6$ one has:
\begin{equation}
\label{eq2-4}
D_1^{(j)}\tau_{\underline\beta}\cdot \tau_{\underline\beta+\underline\delta_i-\underline\delta_k}=
\epsilon_{ijk}\tau_{\underline\beta+\underline\delta_i-\underline\delta_j}\tau_{\underline\beta+\underline\delta_j-\underline\delta_k}
, \quad j=1,2,3,\quad  i,k=1,2,\ldots,6
\, .
\end{equation}
Finally we note that (\ref{equaltau}) can be rewritten to 
\begin{equation}
\label{equaltau2}
\tau_{\underline \alpha}=(-1)^{\alpha_2}\,
\tau_{\underline \alpha+\underline\delta_1+\underline\delta_2+\underline\delta_3-\underline\delta_4-\underline\delta_5-\underline\delta_6}\, .
\end{equation}

\section{From KP to the Jimbo-Miwa-Okamoto $\sigma$-equation} 
To obtain the Jimbo-Miwa-Okamoto $\sigma$-form (\ref{eq:jmo}) of
the Painlev\'e VI equation from the 3-component KP, the following choice of new variables was used in \cite{AL-IMRN} and a similar choice was made in \cite{joshi}:
\begin{equation}
\label{th}
t= \frac{x_1^{(2)}-x_1^{(1)}}{x_1^{(3)}-x_1^{(1)}}\, , \qquad
h=x_1^{(2)}-x_1^{(1)}\, 
\end{equation}
and 
\begin{equation}
\label{difuth}
\pder{}{{x_1^{(1)}}} = \frac{t(t-1)}{h} \pder{}{t}-\pder{}{h},\;\;\; \,
\pder{}{{x_1^{(2)}}}= \frac{t}{h}\pder{}{t}+\pder{}{h},\;\;\; \,
\pder{}{{x_1^{(3)}}}=-\frac{t^2}{h}\pder{}{t}\, .
\end{equation}
Then for $\underline\alpha\in Q(A_5)$ with $R(\underline\alpha)\ge 0$
\[
\pder{{\tau_{\underline \alpha}(t,h)}}{h}=R(\underline\alpha)\tau_{\underline \alpha}(t,h)\, ,
\]
thus 
\[
\tau_{\underline \alpha}(t,h)=h^{R(\underline\alpha)}T_{\underline \alpha}(t)\, .
\]
Using this and equation (\ref{difuth})
equation (\ref{eq1}) turns into (cf \cite{Okamoto}),
\begin{equation}
\label{TodaT}
\begin{split}
R(\underline\alpha)T_{\underline\alpha}^2-(t-1)t^2\left(\frac{dT_{\underline\alpha}}{dt}\right)^2+t^2T_{\underline\alpha}\left(\frac{dT_{\underline\alpha}}{dt}+(t-1)\frac{d^2T_{\underline\alpha}}{dt^2}\right)&=
T_{\underline\alpha+\underline\delta_1-\underline\delta_2}T_{\underline\alpha+\underline\delta_2-\underline\delta_1}\,  , \\
t^2
\left(
t(t-1)
\left(
\frac{dT_{\underline\alpha}}{dt}
\right)^2+
T_{\underline\alpha}
\left(
(1-2t)\frac{dT_{\underline\alpha}}{dt}-t(t-1)\frac{d^2T_{\underline\alpha}}{dt^2}
\right)\right)&=
T_{\underline\alpha+\underline\delta_1-\underline\delta_3}T_{\underline\alpha+\underline\delta_3-\delta_1}\,  ,
\\
t^2\left(-t\left(\frac{dT_{\underline\alpha}}{dt}\right)^2+
T_{\underline\alpha}\left(\frac{dT_{\underline\alpha}}{dt}+t\frac{d^2T_{\underline\alpha}}{dt^2}\right)\right)&=
T_{\underline\alpha+\underline\delta_2-\underline\delta_3}T_{\underline\alpha+\underline\delta_3-\underline\delta_2}\, .
\end{split}
\end{equation}
This gives 3 series of Toda equations that can be used to calculate neighboring tau-functions.
Equation (\ref{eq2-4}) turns into (\ref{blabla0}),
with
\begin{equation}
\label{nnn}
n_1(\underline\alpha;i,k)=-n_2(\underline\alpha;i,k)=R(\underline \alpha+\underline\delta_i-\underline\delta_k)-R(\underline \alpha),\qquad
n_3(\underline\alpha;i,k)=0\, 
\end{equation}
and (\ref{equaltau2}) into 
\begin{equation}
\label{equaltau3}
T_{\underline \alpha}=(-1)^{\alpha_2}\,
T_{\underline \alpha+\underline\delta_1+\underline\delta_2+\underline\delta_3-\underline\delta_4-\underline\delta_5-\underline\delta_6}\, .
\end{equation}
Finally, we have the two Hirota-Miwa equations (\ref{Miwa1}) and (\ref{Miwa2}), that give:
\begin{equation}
\label{MiwaT1}
\begin{split}
&T_{\underline\beta+\underline\delta_2+\underline\delta_3}
T_{\underline\beta+\underline\delta_1+\underline\delta_\ell}-
T_{\underline\beta+\underline\delta_1+\underline\delta_3}
T_{\underline\beta+\underline\delta_2+\underline\delta_\ell}
+
T_{\underline\beta+\underline\delta_1+\underline\delta_2}
T_{\underline\beta+\underline\delta_3+\underline\delta_\ell}
=0,\quad\mbox{for } \ell>3,\quad \mbox{and}\\
&\epsilon_{k\ell}
T_{\underline\beta+\underline\delta_\ell+\underline\delta_i}
T_{\underline\beta+\underline\delta_k+\underline\delta_j}
+\epsilon_{\ell k}
T_{\underline\beta+\underline\delta_k+\underline\delta_i}
T_{\underline\beta+\underline\delta_\ell+\underline\delta_j}
+\epsilon_{j-3,i-3}
T_{\underline\beta+\underline\delta_k+\underline\delta_\ell}
T_{\underline\beta+\underline\delta_i+\underline\delta_j}=0,\quad\mbox{for}
\end{split} 
\end{equation}
$1\le k, \ell\le 3$, $4\le i, j\le 6$ with $i\ne j$ and $k\ne \ell$. All these equations give B\"acklund transformations for the tau functions $T_{\underline\alpha}$ of the Painlev\'e VI equation.

All the above type of equations in the case of the affine Lie algebra of type $A_n$ were obtained by Noumi and Yamada, see e.g. \cite{N} and \cite{NY1}.

We want to rewrite (\ref{blabla0}) and express it in the corresponding Jimbo-Miwa-Okamoto $\sigma$-functions. 
First, we introduce
\begin{equation}
\label{falpha}
f_{\underline\alpha}(t)=t(t-1)\frac{d\log T_{\underline\alpha}}{dt}\, ,
\end{equation}
and take the $\log$ of the expression (\ref{blabla0})
\begin{equation}
\label{blabla2}
\begin{split}
\log&({\rm constant})+\log(T_{\underline\alpha+\underline\delta_i-\underline\delta_j})+\log(T_{\underline\alpha+\underline\delta_j-\underline\delta_k})=\\[2mm] 
&=\log\left(T_{\underline\alpha+\underline\delta_i-\underline\delta_k}\partial_j(T_{\underline\alpha})-T_{\underline\alpha}\partial_j(T_{\underline\alpha+\underline\delta_i-\underline\delta_k})+n_j(\underline\alpha;i,k)T_{\underline\alpha} T_{\underline\alpha+\underline\delta_i-\underline\delta_k}\right)\\[2mm]
&=\log(T_{\underline\alpha+\underline\delta_i-\underline\delta_k})+\log(T_{\underline\alpha})+
\log\left(\frac{\partial_j(T_{\underline\alpha})}{T_{\underline\alpha}}-\frac{\partial_j(T_{\underline\alpha+\underline\delta_i-\underline\delta_k})}{T_{\underline\alpha+\underline\delta_i-\underline\delta_k}}+n_j(\underline\alpha;i,k)\right)\\
&=\log\left(\frac{b_j(t)}{t(t-1)}\right)+\log(T_{\underline\alpha+\underline\delta_i-\underline\delta_k})+\log(T_{\underline\alpha})+\\[2mm]
&\qquad + 
\log\left(\frac{t(t-1)\frac{d}{dt}(T_{\underline\alpha})}{T_{\underline\alpha}}-t(t-1)\frac{\frac{d}{dt}(T_{\underline\alpha+\underline\delta_i-\underline\delta_k})}{T_{\underline\alpha+\underline\delta_i-\underline\delta_k}}+n_j(\underline\alpha;i,k)\frac{t(t-1)}{b_j(t)}\right)\\[2mm]
&=\log\left(\frac{b_j(t)}{t(t-1)}\right)+\log(T_{\underline\alpha+\underline\delta_i-\underline\delta_k})+\log(T_{\underline\alpha})+\\[2mm]
&\qquad + 
\log\left(t(t-1)\frac{d\log(T_{\underline\alpha})}{dt}-t(t-1)\frac{d\log(T_{\underline\alpha+\underline\delta_i-\underline\delta_k})}{dt}+n_j(\underline\alpha;i,k)\frac{t(t-1)}{b_j(t)}\right)\, .
\end{split}
\end{equation}
Now take $t(t-1)\frac{d}{dt}$ of this expression (\ref{blabla2}), we thus obtain:
\begin{equation}
\label{blabla3}
\begin{split}
f_{\underline\alpha+\underline\delta_i-\underline\delta_j}(t)&+f_{\underline\alpha+\underline\delta_j-\underline\delta_k}(t)-f_{\underline\alpha+\underline\delta_i-\underline\delta_k}(t)-f_{\underline\alpha}(t)=\\
&=g_j(t)
+t(t-1)\frac{d}{dt}
\log\left(f_{\underline\alpha}(t)
-f_{\underline\alpha+\underline\delta_i-\underline\delta_k}(t)+h_j(t) \right)\, ,
\end{split}
\end{equation}
where
\begin{equation}
\label{gh}
g_j(t)=
\begin{cases}0&{\rm if}\qquad   $j=1$,\\
-t&{\rm if}\qquad  $j=2$,\\
1&{\rm if}\qquad   $j=3$,\\
\end{cases}\qquad
h_j(t)=
\begin{cases}
n_j(\underline\alpha;i,k)&{\rm if}\qquad   $j=1$,\\
n_j(\underline\alpha;i,k)(t-1)&{\rm if} \qquad  $j=2$,\\
-n_j(\underline\alpha;i,k)\frac{t-1}{t}=0&{\rm if}\qquad   $j=3$.\\
\end{cases}\qquad
\end{equation}
Following \cite{AL-IMRN} we introduce 
\begin{equation}
\label{sigma-alpha}
\sigma_{\underline\alpha}=f_{\underline\alpha}(t) +c_5(\underline\alpha)(t-1)-\frac12c_6(\underline\alpha)\, ,
\end{equation}
where
\begin{equation}
\label{c5c6}
\begin{split}
c_5(\underline\alpha)&=-\frac14( \alpha_1-\alpha_3)^2\, ,
\\[2mm]
c_6(\underline\alpha)&=R(\underline\alpha)+\frac12(\alpha_1-\alpha_2)( \alpha_1-\alpha_3)\, ,
\end{split}
\end{equation} 
and thus we obtain (\ref{blabla40}),
where
\begin{equation}
\label{GH}
\begin{split}
G_{ijk}(\underline\alpha;t)=&\left(c_5({\underline\alpha+\underline\delta_i-\underline\delta_j})+c_5({\underline\alpha+\underline\delta_j-\underline\delta_k})-c_5({\underline\alpha+\underline\delta_i-\underline\delta_k})-c_5({\underline\alpha})\right)(1-t)\\&+\frac12\left(c_6({\underline\alpha+\underline\delta_i-\underline\delta_j})+c_6({\underline\alpha+\underline\delta_j-\underline\delta_k})-c_6({\underline\alpha+\underline\delta_i-\underline\delta_k})-c_6({\underline\alpha})\right)+g_j(t)\, ,
\\[2mm]
H_{ijk}(\underline\alpha;t)=&\left(c_5({\underline\alpha})-c_5({\underline\alpha+\underline\delta_i-\underline\delta_k})\right(1-t)+\frac12\left(c_6({\underline\alpha})-c_6({\underline\alpha+\underline\delta_i-\underline\delta_k})\right)+h_j(t)\, .
\end{split}
\end{equation}

\section{Other B\"acklund transformations}

Besides the  B\"acklund transformations that come from the 3-component Grassmannian structure, there are some other relevant transformations.
A first observation that can be made is that the  $\sigma$ equation 
(\ref{eq:jmo}) has a natural $D_4$ symmetry. One can permute all $v_i$ together with an even number of sign changes.

Secondly, one can choose an other identification (\ref{v-in-alpha-mu0}) between the $\alpha$'s and $v$'s see e.g. \cite{AL-IMRN}, section 2.

Thirdly, one can permute the $\alpha_i$'s for $i=1,2,3$ and also separately the $\mu_i$'s.
All these transformations rearrange the tau functions on the $sl_6$ root lattice.

Finally, starting with the underlying 3-component KP model
one can choose a different identification (\ref{th}) between
the $x_1^{(i)}$ and $t$ and $h$. For instance interchanging
$x_1^{(1)}$ and $x_1^{(3)}$, gives a transformation $t\mapsto 1-t$,
such a transformation leaves Painlev\'e VI equation (\ref{eq:P6})
invariant for  $y\mapsto 1-y$ and appropriate transformations of
coefficients, see e.g. Boalch \cite{Boalch} or \cite{NY2}.
The permutation that interchanges $x_1^{(1)}$ and $x_1^{(2)}$
(respectively $x_1^{(2)}$ and $x_1^{(3)}$), gives a transformation
$t\mapsto \frac{t}{t-1}$ (resp. $t\mapsto \frac{1}{t}$), such a
transformation induces
$y\mapsto\frac{t-y}{t-1}$ (resp. $y\mapsto \frac{1}{y}$), again see \cite{Boalch} or \cite{NY2}, where it is argued that addition of
these transformations extend $D_4^{(1)}$  symmetry to  $F_4^{(1)}$ symmetry.

\section{Root lattice of $F_4^{(1)}$}

Okamoto showed in his fundamental paper \cite{Okamoto} that there is a representation of the affine Weyl group of type 
$F_4^{(1)}$ that acts on the solutions of the Painlev\'e VI equation. An element in this Weyl group is related to a birational canonical transformation.
We will now show that the $sl_6$ root lattice of the previous section is related to the root lattice of the affine Lie algebra of type $F_4^{(1)}$ on which this affine Weyl group acts.

Let 
\begin{equation}
\label{vvv}
\underline v=(v_0,v_1,v_2,v_3,v_4)=v_0\underline e_0+v_1\underline e_1+v_2\underline e_2+ v_3\underline e_3+v_4\underline e_4
\end{equation}
be a vector in a 5-dimensional vector space. We assume that 
\[
(\underline v,\underline w)=\sum_{i=1}^4 v_iw_i\, .
\]
If  we make the following identification (see also (\ref{v-in-alpha-mu0})):
\begin{equation}
\label{v-in-alpha-mu}
v_0=\alpha_1,\quad
v_i=\frac{\alpha_1+\alpha_3}2+\mu_i=\frac{\alpha_1+\alpha_3}2+\alpha_{3+i}\quad (i=1,2,3),\quad v_4=\frac{\alpha_1-\alpha_3}2,
\end{equation}
then the $v_1,\, v_2,\, v_3,\,  v_4$ correspond to the parameters of the Jimbo-Miwa-Okamoto $\sigma$-equation  (\ref{eq:jmo}). Moreover, one  has the following correspondence, the element
$
\sum_{i=1}^6 \alpha_i\underline\delta_i$
is equal to
\[
\alpha_1 \underline e_0+\left(\frac{\alpha_1+\alpha_3}2+\alpha_4 \right)\underline e_1 +
\left(\frac{\alpha_1+\alpha_3}2+\alpha_5 \right)\underline e_2 +
\left(\frac{\alpha_1+\alpha_3}2+\alpha_6 \right)\underline e_3 +
\left( \frac{\alpha_1-\alpha_3}2\right)\underline e_4\, .
\]
Note that $\underline e_0=\underline\delta_1+\underline\delta_2+\underline\delta_3
-\underline\delta_4-\underline\delta_5-\underline\delta_6$.
In this way one gets all elements of the form (\ref{vvv}) with $v_0\in\mathbb{Z}$ and all
$v_i\in\mathbb{Z}$ for $i>0$ or all $v_i\in\frac12+\mathbb{Z}$ for $i>0$. This is the root lattice $Q(F_4^{(1)})$ of the Lie algebra of type $F_4^{(1)}$.
In fact the simple roots of this affine Lie algebra  are:
\[
\begin{split}
\underline e_0-\underline e_1-\underline e_2=&
\underline\delta_1+3\underline\delta_2+\underline\delta_3-2\underline\delta_4
-2\underline\delta_5-\underline\delta_6,\\
\underline e_2-\underline e_3=&
\underline\delta_5-\underline\delta_6,\\
\underline e_3-\underline e_4=&
2\underline\delta_3-\underline\delta_4-\underline\delta_5,\\
\underline e_4=&
-\underline\delta_2-2\underline\delta_3+\underline\delta_4+\underline\delta_5+\underline\delta_6,\\
\frac12(\underline e_1-\underline e_2-\underline e_3-\underline e_4)=&
\underline\delta_2+\underline\delta_3-\underline\delta_5-\underline\delta_6\, .
\end{split}
\]
The $\pm(\underline\delta_i-\underline\delta_j)$ with $1\le i\le 6$, $1\le j\le 3$ and $i\ne j$ that appear in the sigma functions of equation (\ref{blabla40}) form up to possibly a translation with the vector $\underline e_0$ all short roots of $F_4$, which are ($\epsilon_k=\pm 1$):
\[ \epsilon_k\underline e_k,\quad(k=1,2,3,2),\quad \frac12(\epsilon_1\underline e_1+\epsilon_22\underline e_2+\epsilon_3\underline e_3+\epsilon_4\underline e_4)\, .
\]
To be more precise they form the union of the sets $\pm S_j$, wich are  defined by
\begin{equation}
\begin{split}
S_1=&\{
\underline e_0+\underline e_4,
\underline e_0+\frac12(\underline e_1+\underline e_2+\underline e_3+\underline e_4),
-\underline e_0+\frac12(\underline e_1-\underline e_2-\underline e_3+\underline e_4),\\
&\qquad\underline e_0+\frac12(\underline e_1-\underline e_2+\underline e_3+\underline e_4),
\underline e_0+\frac12\underline e_1+\underline e_2-\underline e_3+\underline e_4)
\},\\
S_2=&\{
\underline e_0+\frac12(\underline e_1+\underline e_2+\underline e_3+\underline e_4),
\frac12(\underline e_1+\underline e_2+\underline e_3-\underline e_4),
e_1, e_2, e_3
\},\\
S_3=&\{
\underline e_0+\underline e_4,
\frac12(-\underline e_1-\underline e_2-\underline e_3+\underline e_4),
\frac12(\underline e_1-\underline e_2-\underline e_3+\underline e_4),\\
&\qquad
\frac12(-\underline e_1+\underline e_2-\underline e_3+\underline e_4),
\frac12(-\underline e_1-\underline e_2+\underline e_3+\underline e_4)
\}\, .
\end{split}
\end{equation}
Then the following holds:\\
\ 
\\
{\it
Let $\underline\beta$ be an element in the root lattice of $F_4^{(1)}$ and asume $\gamma_1, \gamma_2\in S_j$ for fixed $j=1,2,3$, suppose $\sigma_{\underline\beta}$ and 
$\sigma_{\underline\beta+\underline\gamma_1-\underline\gamma_2}$ are known, then using equation (\ref{blabla40}) one can calculate 
$\sigma_{\underline\beta+\underline\gamma_1}$ (resp. $\sigma_{\underline\beta-\underline\gamma_2}$), if one knows
$\sigma_{\underline\beta-\underline\gamma_2}$ (resp.  $\sigma_{\underline\beta+\underline\gamma_1}$).
}
\\
\ 
\\
Clearly a similar implication also holds for the corresponding tau functions $T_{\underline\beta}$.
\\
Equation (\ref{TodaT}) implies:
\\
\ 
\\
{\it Let $\underline\beta$ be an element in the root lattice of $F_4^{(1)}$ and asume $\underline\gamma=
\underline e_0+\frac12(\underline e_1+\underline e_2+\underline e_3+\underline e_4)$, 
$\frac12(\underline e_1+\underline e_2+\underline e_3-\underline e_4)$ or
$\underline e_0+\underline e_4$, suppose  $T_{\underline\beta}$ and $T_{\underline\beta\pm \underline\gamma}$ are known, using equation  (\ref{TodaT}) one can calculate  $T_{\underline\beta\mp \underline\gamma}$.
}
\\
\ 
\\
Equation (\ref{equaltau3}) implies:
\\
\
\\
{\it Let $\underline\beta$ be an element in the root lattice of $F_4^{(1)}$
then up to a sign $T_{\underline\beta}$ is equal to  $T_{\underline\beta+\underline e_0}$
}
\\
\
\\
Finally the Hirota-Miwa equation (\ref{MiwaT1}) also gives a connection between six tau functions in the $F_4^{(1)}$ root lattice. However it is not so easy to describe this explicitly in this $F_4^{(1)}$ setting.

\end{document}